# An Internet Framework to Bring Coherence between WAP and HTTP Ensuring Better Mobile Internet Security

Al-Mukaddim Khan Pathan, Md. Abdul Mottalib and Minhaz Fahim Zibran*
Department of Computer Science and Information Technology, Islamic University of Technology, Bangladesh.
* Department of Mathematics and Computer Science, University of Lethbridge, Alberta, Canada
E-mail: mukaddim@iut-dhaka.edu, mamottalib@iut-dhaka.edu, zibranm@yahoo.com

*Abstract* — **To bring coherence between Wireless Access Protocol (WAP) and Hyper Text Transfer Protocol (HTTP), in this paper, we have proposed an enhanced Internet framework, which incorporates a new markup language and a browser compatible with both of the access control protocols. This Markup Language and the browser enables co-existence of both Hyper Text Markup Language (HTML) and Wireless Markup Language (WML) contents in a single source file, whereas the browser incorporates the ability to hold contents compliant with both HTTP and WAP. The proposed framework also bridges the security gap that is present in the existing mobile Internet framework.**

## 1. Introduction

To dates, Internet (World Wide Web) has become a part and parcel of modern life. HTTP (Hypertext Transfer Protocol) was one of the first World Wide Web (WWW) technologies. Today it has been the established and omni-accepted protocol since 1990 [1]. HTTP is now widely used as a distributed application-level or wired-level communication protocol. Web based applications incorporating web browsers and web servers rely on the HTTP for the client-server communication over the wired Internet.

With the rapid progress of technological development the world is moving towards pervasive computing. Computational devices are becoming smaller and lighter in weight. People all over the world are coming forward to use small wireless devices like Personal Digital Assistances (PDAs), palm-top computers and mobile phones. To enable wireless communication among these small wireless devices the new protocol Wireless Application Protocol (WAP) came in. WAP compatible applications are used to carry out the wireless communication.

A person with a desktop PC (Personal Computer) connected with the Internet reads the Hypertext Markup Language (HTML) content of a web page using some HTML-browser (web-browser; e.g., Microsoft Internet Explorer, Netscape Navigator, etc.). But a small wireless device uses a WML-browser (micro-browser; e.g., WinWap, Klondike WAP Browser, etc.) to render the Wireless Markup Language (WML) content of a WML page. The web browser cannot render the contents of WML page and in the same way the micro-browser cannot render the contents of an HTML page.

Thus the knowledge domain of World Wide Web is partitioned into two separate subsets: wired domain and wireless domain. But it would have been evidently better if the HTML pages could be readable using small handheld devices and the WML contents could be rendered by the PC web-browsers. To make this facility available, several endeavors have been taken in the software industry and by the research groups. But the success rate and accuracy of those efforts are not sufficient. Moreover, some security pitfalls are still present in the present WAP implementation framework. In order to address the above mentioned issues, in this paper, a new Internet framework (i.e., HTTP-WAP framework) has been proposed, which overcomes the security gap in the existing mobile Internet and brings coherence between WAP and HTTP using a unified markup language and an HTML-WML browser.

The rest of the paper is organized as follows: Section 2 describes the existing access control protocols for wired and wireless communication, Section 3 depicts the security gap that is present in the existing mobile Internet framework, Section 4 discusses the issues to bring coherence between WAP and HTTP and covers the review of some related works, Section 5 presents the proposed Internet framework for HTTP and WAP, Section 6 deals with the performance evaluation and Section 7 concludes the paper.

## 2. Existing Access Control Protocols for Wired and Wireless Communication

Hyper Text Transfer Protocol (HTTP) is now widely used as a distributed application-level or wired-level communication protocol; whereas Wireless Access Protocol (WAP) enables wireless communication among the small wireless hand-held devices. WAP compatible applications are used to carry out the wireless communication. These two existing access control protocols are described below:

### 2.1. Hyper Text Transfer Protocol (HTTP)

The Hyper Text Transfer Protocol (HTTP) is an application-level protocol for distributed, collaborative,

hypermedia information systems [2]. HTTP has been in use by the World-Wide Web global information initiative since 1990. "HTTP/1.1" is an update to RFC 2068 [3]. HTTP, which follows the request/response paradigm usually, takes over TCP/IP (Transmission Control Protocol / Internet Protocol); however, HTTP is not dependent on TCP/IP. In the client-server communication chain intermediates like gateways, tunnels or proxies can be present [1], as shown in figure 1.

HTTP supports a large verity of content types, such as plain text, HTML, images (BMP, JPEG, GIF, etc.) movie files (MPEG, MOV, etc.) and many others.

2.2. Wireless Application Protocol (WAP)

HTTP in collaboration with other internetworking protocols works pretty good in networks connected with some guided media. But for wireless communication bandwidth is comparatively smaller. And underlying small mobile devices have some limitations in processing ability, memory, I/O and power consumption. To cope with these limitations WAP came in.

WAP is a protocol for accessing information and services from wireless devices. In parallel to the efforts to establish i-mode in Japan, in June 1997, Ericson, Motorola, Nokia and Openwave (formerly known as Unwired Planet and Phone.com) founded WAP forum as an industry group for the purpose of extending Internet standards for the use with the wireless communication.

*2.2.1. WAP communication model*

WAP is used for micro browsers in wireless devices whereas HTTP is used for web browsers and mostly for wired Internet – it allows them to become clients in an Internet-based client/server world. Micro-browsers in WAP devices connect to servers to retrieve and send information in much the same way as web-browsers in PCs connect to HTTP servers. In order to serve WAP content one has to install the WAP server. This piece of software is much like an HTTP server and indeed, the two can usually run on the same machine. WAP devices can directly connect WAP servers [4] as shown in figure 2.

Almost all Mobile Internet infrastructure implementation uses an intermediary WAP Gateway between the WAP Client and HTTP Server, as shown in figure 3. WAP protocols are used between WAP client and the WAP gateway. On the other hand, TCP/IP and HTTP are used between the WAP gateway and HTTP server [1]. It is the responsibility of the WAP gateway to translate requests from WAP stack to WWW protocol stack (TCP/IP and HTTP) and to decode requests sent from the WAP client to server. The responses from the server are encoded by the gateway into a compact binary format, which the client is able to interpret.

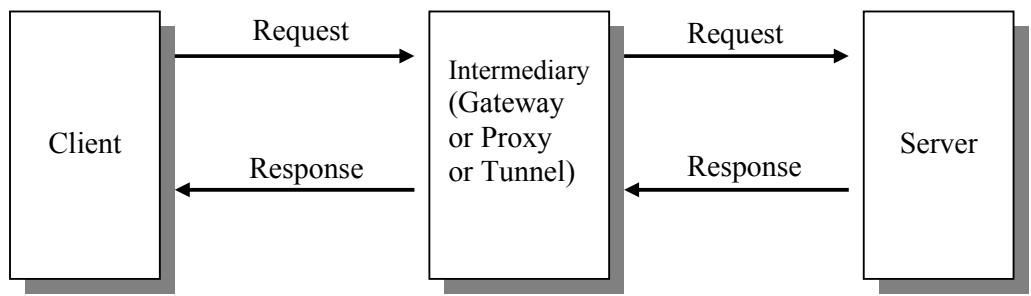

**Figure 1. Use of intermediary in HTTP communication Model**

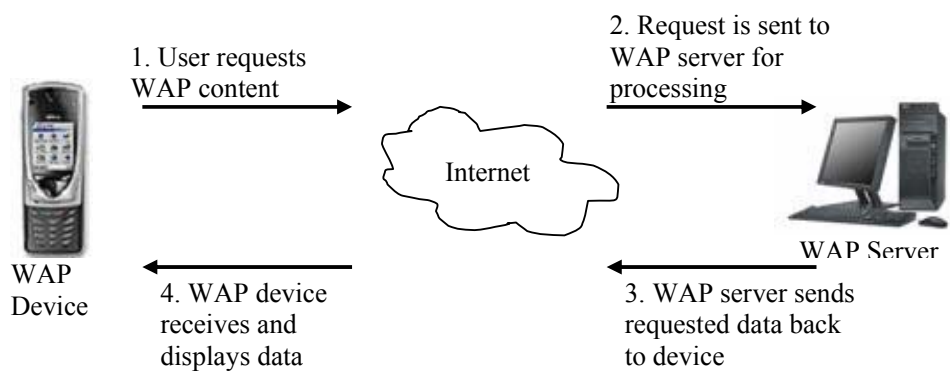

**Figure 2. Direct communication between WAP Device and WAP Server**

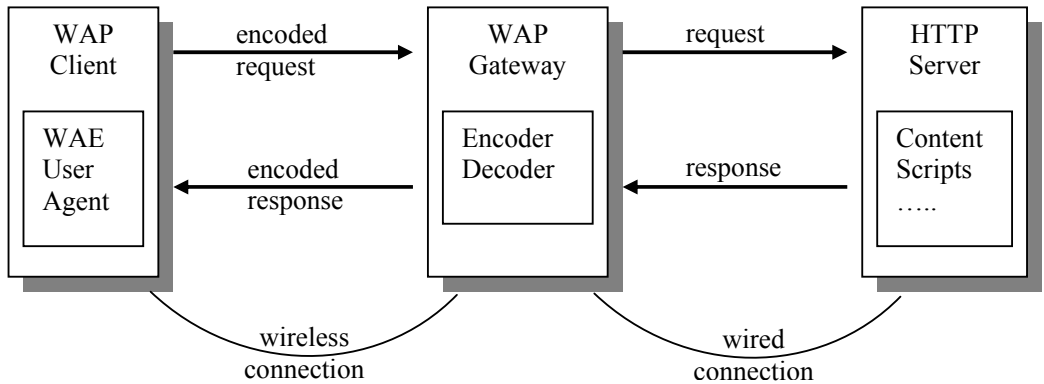

Figure 3. WAP infrastructure using intermediate gateway

## 3. The Security Gap [5]

If security is required in the fixed-wire world, Transport Layer Security (TLS) is usually used, and in the wireless world there is the Wireless Transport Layer Security (WTLS). As mentioned in section 2.2.1, the WAP client directly communicates with the WAP gateway. WTLS is the security protocol that is used to secure communications to and from the mobile device, but the mobile device's session is necessarily with the WAP gateway rather than the remote host's web server. The WAP gateway uses TLS to establish a secure session for secured communication with the server. It is the gateway, which performs the necessary encoding and decoding, as stated in section 2.2.1.

There are actually two secure sessions in play: one between the mobile device and the WAP gateway and the other between the WAP gateway and the web server. This means that there is a security gap, in which the data are not encrypted, at the WAP gateway, as shown in figure 4. Hence, it can become the target of hackers. So, this security gap should be eliminated by taking necessary measures.

## 4. Bringing Coherence between Wired and Wireless Internet

As mentioned before, a person with a PC using wired Internet has a web-browser (Internet Explorer, Opera, etc.), with which he sees HTML contents (and other HTTP supported resources) of some web site. This PC user cannot see the WML contents (and other WAP supported resources) using the web-browser of the PC. To connect to some web site he/she uses address like this: www.somesitename.com. This connection is done by the HTTP. Although www is the de facto standard hostname for HTTP servers, WAP seems to be emerging as a comparable standard for servers containing WAP applications. WAP sites use the naming convention like: wap.somesitename.com.

Thus, only the WAP sites (WML contents) are available for browsing to the people using mobile devices. On the contrary, only the web sites (HTML contents), not the WAP sites are available to the PC users. This brings some negative consequences-

- Information of the WAP sites is accessible only to the mobile device users, whereas, information of web sites is accessible only to the PC users.
- An organization has to launch and maintain two different sites with redundant information, one WAP site for the mobile device users and another web site for the PC users.
- Programmers, designers and developers, while developing sites, have to target either of the PC web-browser and micro-browser of mobile devices and not the both.
- Eventually, the Internet knowledge domain is partitioned into two sub-domains: wired Internet and wireless Internet.

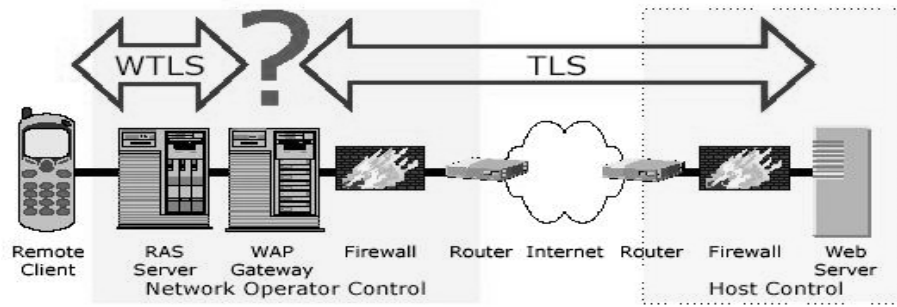

Figure 4. Security gap in present WAP framework

### 4.1. Related works

To prevent the evolving logical partition between the information domains of wired and wireless Internet, researchers and Internet professionals are still looking for a suitable solution. Almost all of the previous works were attempts to develop HTML-to-WML translators. Such translators are used to translate HTML or XML content of web sites to WML content, in order to be rendered by the micro browser. The objective was ultimately to provide mobile Internet users with a useful tool for web access.

Kapadia [6] proposed three different solutions to convert XML content to WML. First one is to write a Java program that reads the input, extracts the required data, adds WML tags where appropriate and outputs a .wml file. The second solution involves using XSLT, the XML parser Xerces [7], the XSLT processor Xalan [7] and a Java file to apply the conversion. The third method is based on Java Servlets that use Cocoon [8] and works like a web server only responding to URL requests by publishing files transformed as specified. Though he proposed some interesting designs, the scope is rather too limited since input was from known XML structures. Again, the software would have to be re-written in order to deal with new tags and HTML content was not considered in any of the designs. Html2Wml Version 0.4.1 [9] is a HTML to WML conversion tool which converts HTML content to WML on the fly. But, input to the program must be valid well-formed HTML, whilst the output is far from valid WML and so incapable of immediately rendering on a WML browser. Moreover, there is no provision for support of frames. LazyWAP [10] is a HTML to WML converter which on the fly converts HTML contents to WAP compatible format. Here the input HTML files should be XHTML compliant. But it has no routines for handling anything more than very simple input and the involvement of string replacement highlights the difficulty and sloppiness of its implementations in Java, C or PHP. Translation of XML to HTML can be done through eXtensible Stylesheet Language Transformations (XSLT) [11]. But this implementation is again restricted to a known type of XML and XSLT is used over other languages. Kaasinen et al. proposed methods for handling frames and complex HTML conversion [12]. Their delivery of HTML tables to the small output screens jumbled up the sorting of links items and confused the users. Again, converting malformed HTML has placed restrictions upon their software.

### 4.2. Potential challenges

Though a number of initiatives have been taken to bring coherence between HTTP and WAP, unfortunately their success rate and accuracy are very small. In order to address this situation, a new Internet framework (i.e., HTTP-WAP framework) is necessary which may incorporate the conversion between WAP content and conventional HTTP content. This conversion involves the following three challenges:
- Translation between HTML and WML.
- Translation between WMLScript and web-scripting languages like JavaScript, VBScript, etc.
- Transformation between Wireless Application Environment (WAE) supported content type (WBMP, for example) and normal web content (BMP, JPEG, etc., for example)

*4.2.1. Translation between HTML and WML*

HTML has well over 120 tags [13]. HTML rule are defined in Standard Generalized Markup Language (SGML), which is an international publishing standard in existence since 1986. XML (eXtensible Markup Language) is a successor to SGML and WML is an XML application (i.e., XML defined language) [14]. WML has 35 strictly applied semantic tags intended for delivery using the WAP [15]. WML tags are in some way logically subset of HTML tags. Therefore, it is not always possible to substitute an HTML tag of an HTML file with a corresponding WML tag [16].

Since WML is an XML application, it is well-formed. But HTML is not well-formed. Rules for HTML tag nesting are loose and not always enforced by the web-browsers. At the very beginning of the translation process, the underlying HTML page has to be converted into well-formed HTML (i.e., XHTML) [17].

HTML tags can be both uppercase and lowercase. On the contrary, WML tags must all be in lowercase. Therefore, all HTML tags have to be converted to lowercase.

Again, the organization of a WML document largely varies from that of an HTML document. An HTML file contains a single HTML page. But a WML file may contains a single WML deck, and multiple WML cards constitute a single WML deck. Viewing an HTML page is actually viewing the content of an HTML file, whereas in case of WML a single card of a WML deck is displayed at a time. Hence, for translation between HTML and WML this difference in content organization is to be handled properly.

*4.2.2. Translation of Scripting Language*

WMLScript is a client side scripting language to be used with WAP. One major difference of WMLScript with conventional web scripting languages (i.e., JavaScript, VBScript, etc.) is the use of WMLScript byte code and byte code interpreter. The WAP gateway compiles the WMLScript into byte code before transmitting it to the client [1]. WAP also defines a micro Virtual Machine (VM) for use by the micro-browser to execute the WMLScript byte code. Translation between HTML and WML also necessitates translation between WMLScript and conventional web scripting languages.

*4.2.3. Transformation between WAE Content and HTTP Content*

WAP defines the WAE content types, which are suited to the limited memory and CPU constraints of mobile devices. Translation between WML and HTML also requires transformation of WAE content into HTTP web content and vice versa.

## 5. The Proposed Internet Framework for HTTP and WAP

We have proposed some modifications both in WAE specification and WAP framework itself, which include:
- Replacement of WML with wHTML which is compatible with both WAP and HTTP.
- File implementation of WMLScript for allowing byte code reuse.
- A universal browser capable of rendering both HTTP and WAP content.
- Guideline for secured WAP implementation.

### 5.1. Markup Language compatible with both WAP and HTTP

We have introduced a markup language which is compatible with both WAP and HTTP. We have named the proposed Markup Language as 'wHTML' (Wireless Hypertext Markup Language). According to the XML like specification, wHTML will be well-formed. The source file containing the wHTML content will have the .wHTML extension. This wHTML language constitutes a set of tags, which is the union of both HTML tag set and WML tag set. Within the wHTML tag set HTML tags will have 'h' prefix, whereas WML tags will have 'w' prefix. This prefix convention will enable the parser to differentiate between HTML tags and WML tags.

### 5.2. File implementation of WMLScript

The WMLScript codes to be used with a WML (for the proposed framework, it is wHTML) document would be written in separate file. The implementation would be similar to the byte code generation of Java. Before deploying a wHTML site in the server, the developer would compile the source file containing WMLScript source code and generate corresponding file consisting the WMLScript byte code. For referencing the functional code the wHTML document will just refer to the file containing the byte code. This will allow byte code reusability and free the gateway from generating the WMLScript byte code every time a page is requested.

### 5.3. A universal browser Capable of rendering both HTTP and WAP content

Usually web-browsers and micro-browsers use parsers of different types with distinct parsing logics. We have proposed a browser, which will use a wHTML parser. This parser will check whether the wHTML document is well-formed according to specification. This checking is done in a single pass over the document, using a stack [14]:
  a. As the parser encounters a start tag, pushes it on the stack.
  b. As it encounters a matching end tag, it pops the start tag off the stack.
  c. If the stack is empty at the end then the document is well-formed.
  d. If any time during a pass, the parser comes across an end tag that does not match with the start tag at the top of the stack, the document is not well-formed.
  e. If the parser reaches the end of the document and there are still some start tags left on the stack, the document is not well-formed.
  f. A wHTML document may have tags with either 'h' or 'w' prefixes or even no prefix at all. While preparing the parse tree the parser knows whether it has to prepare a parse tree for WAP or HTTP. For WAP, it simply ignores the tags with 'h' prefix and for HTTP it ignores tags with 'w' prefix. In accordance to the generated parse tree the parser renders the content of the page to be displayed, be it for WAP or HTTP.

### 5.4. Secured WAP-HTTP implementation guideline

To visit some www site, a user uses URL like http://www.somesitename.com. However, for secured transaction the user uses https://www.somesitename.com. According to our proposal, for visiting a WAP site, the user will use URL like wap://www.somesitename.com, where for secured communication he/she will use waps://www.somesitename.com. Thus it will bridge the security that is incurred in the existing WAP framework.

The proposed Internet framework for HTTP and WAP is shown in figure 5:

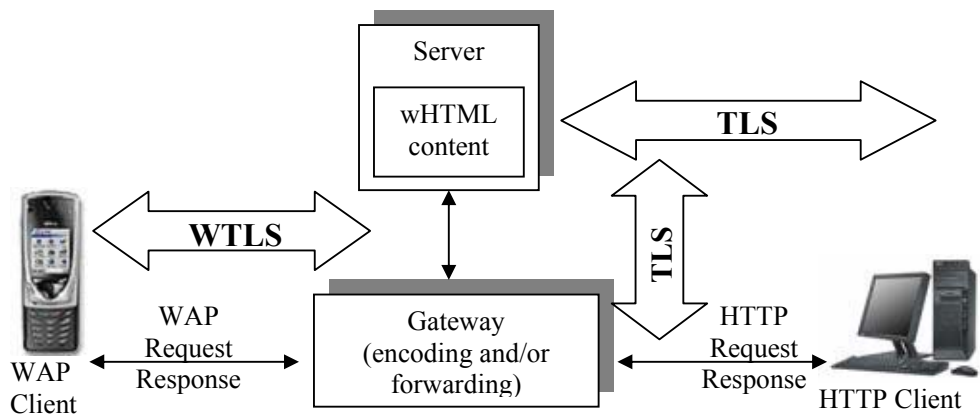

**Figure 5. The Proposed Internet Framework for HTTP and WAP**

This communication framework has a gateway between the client and the server. Whenever the server receives an HTTP or WAP request, in response it sends back the requested wHTML content to the gateway. The gateway parses the wHTML document, filters out the HTML or WML content and sends this back either to the HTTP client or to the WAP client who has made the request. The security of the communication between the server and the gateway is ensured by TLS. TLS also provides secure communication between the HTTP client and the gateway, while WTLS ensures secured communication between the WAP client and the gateway. Upon receiving the content the browser renders the display.

When the server receives an HTTPS (HTTP Secured) or WAPS (WAP Secured) request, it sends back the requested wHTML content to the gateway and the gateway just forwards it to the requesting client. Security of the whole communication is assured by the SSL (Secured Session/Socket Layer). Upon receiving the wHTML, the browser parses and sorts out the HTTP or WAP content based on the request type and renders the display.

Here, in case of WAPS communication data are never in decrypted form in their way, not even in the gateway. So the security gap mentioned in section 3 does not exist in the proposed Internet framework.

## 6. Performance Evaluation

It has been found that the proposed Internet framework has the following advantages over the existing one:

- The most significant advantage of the proposed framework is unifying the wired and wireless Internet information domains and bringing them in a single platform.
- The proposed framework fills up the security gap in the existing mobile Internet framework by compromising network traffic only at the time of secured transaction (HTTPS or WAPS).
- During secured transaction, gateway just forwards the data. Thus, it incurs reduced overhead.
- Network traffic does not increase at all in case of normal Internet communication (HTTP and WAP), though in such case parsing overhead in the underlying browser rises a bit.
- In the proposed framework, file implementation of WMLScript allows the byte code reusability and free the gateway from generating the WMLScript byte code every time a page is requested.
- wHTML is supposed to have XML like specification to be well-formed. So, in the proposed framework, the parser checks whether the wHTML document is well-formed. Thus, the wHTML specification for *"well-formed Markup Language"* is met.

## 7. Conclusion

In the near future WAP is likely to get out of the constraints of scarce memory, bandwidth and processing ability. The divergences between WAP and HTTP will become negligible in course of time. To bring coherence between WAP and HTTP, in the proposed Internet framework, a new markup language and a browser compatible with both of the access control protocols are incorporated. This Markup Language and the browser enables co-existence of both HTML and WML contents in a single source file, whereas the browser incorporates the ability to hold contents compliant with both HTTP and WAP. The proposed Internet framework for HTTP and WAP appears to be a good solution to keep the Internet knowledge domain unified and centralized. The proposed framework also bridges the security gap that is present in the existing mobile Internet framework. Thus HTTP and WAP are brought in a common platform in spite of the divergence that exists between them and better mobile Internet security is ensured. Therefore, we hope that the proposed Internet framework will be a significant contribution to the on-going revolution of Internet technology.